\documentclass[fleqn,usenatbib]{mnras}

\usepackage[T1]{fontenc}

\DeclareRobustCommand{\VAN}[3]{#2}
\let\VANthebibliography\thebibliography
\def\thebibliography{\DeclareRobustCommand{\VAN}[3]{##3}\VANthebibliography}

\usepackage{graphicx}	% Including figure files
\usepackage{graphicx,times}
\usepackage{amsmath}	% Advanced maths commands
\usepackage{amssymb}	% Extra maths symbols
\usepackage{amssymb,amsmath}
\usepackage[]{hyperref}

\title{Evidence of Dark Contents in the Center of NGC 6517}

\author[Yi Xie]{
Yi Xie $^{1}$\thanks{E-mail: xieyi@jmu.edu.cn}
Dejiang Yin$^{2}$
Lichun Wang$^{1}$
Yujie Lian$^{3,4}$
Liyun Zhang$^{2}$
Zhichen Pan$^{5}$
\\
$^{1}$School of Science, Jimei University, Xiamen 361021, Fujian Province, China\\
$^{2}$College of Physics, Guizhou University, Guiyang 550025, China\\
$^{3}$Institute for Frontiers in Astronomy and Astrophysics, Beijing Normal University, Beijing 102206, China\\
$^{4}$Department of Astronomy, Beijing Normal University, Beijing 100875, China\\
$^{5}$National Astronomical Observatories, Chinese Academy of Sciences, Beijing 100101, China\\
}

\date{Accepted XXX. Received YYY; in original form ZZZ}

\pubyear{2022}

\begin{document}
\label{firstpage}
\pagerange{\pageref{firstpage}--\pageref{lastpage}}
\maketitle


\begin{abstract}
Millisecond pulsars can serve as effective probes to investigate the presence of Intermediate-mass Black Holes (IMBHs) within Galactic globular clusters (GCs). Based on the standard structure models for GCs, we conduct simulations to analyze the distributions of pulsar accelerations within the central region of NGC 6517. By comparing the measured accelerations of pulsars obtained from their period derivatives $\dot P$ to the simulated distribution profiles, we demonstrate that a central excess of dark mass is required to account for the measured accelerations. Our analysis, which relies on existing pulsar timing observations, is currently unable to differentiate between two possible scenarios: an IMBH precisely situated at the core of the cluster with mass $\gtrsim 9000^{+4000}_{-3000}~M_{\odot}$, or a central concentration of stellar mass dark remnants with a comparable total mass. However, with additional acceleration measurements from a few more pulsars in the cluster, it will be possible to differentiate the source of the nonluminous matter.

\end{abstract}

\begin{keywords}
(stars:) pulsars: general -- (Galaxy:) globular clusters: general -- stars: black holes
\end{keywords}

%%%%%%%%%%%%%%%%%%%%%%%%%%%%%%%%%%%%%%%%%%%%%%%%%%

%%%%%%%%%%%%%%%%% BODY OF PAPER %%%%%%%%%%%%%%%%%%

\section{Introduction}
Intermediate-mass black holes (IMBHs), possessing masses within the range of $10^2$ and $10^5 M_{\odot}$, are the critical bridge to connect stellar-mass black holes (BHs) and supermassive BHs. Although conclusive evidence for the existence of IMBHs is still lacking, they are proposed to be situated within dwarf galaxies \citep{2004ApJ...610..722G,2015ApJ...809L..14B}, nearby globular clusters (GCs) \citep{2004Natur.428..724P,2017Natur.542..203K}, or appearing as ultraluminous X-ray sources \citep{2001ApJ...551L..27M,2009Natur.460...73F}. The existence of IMBHs within GCs has been extensively investigated for a long time \citep{1975ApJ...200L.131S,1975Natur.256...23B}. Theoretical models suggested that IMBHs could be formed as a result of runaway core collapse and subsequent merger of stars within highly dense star clusters \citep{2002ApJ...576..899P}. Therefore, the central regions of GCs present an ideal environment for the formation of IMBHs, given the high frequency of stellar collisions and mergers \citep{2002MNRAS.330..232C,2004ApJ...604..632G}.

Three types of evidence have been widely invoked to demonstrate the existence of IMBHs in GCs. Firstly, surface photometry for the Galactic GCs discovered that about one fifth of the known GCs display indications of a collapsed core \citep{1986ApJ...305L..61D}. In other word, their surface brightness profiles exhibit a central power-law cusp. A potential interpretation could be that the IMBHs might reside at the core of cusp clusters \citep{1976ApJ...208L..55N,2008ApJ...676.1008N}. Through optical observations, one can also measure various aspects of projected density and kinematical data, including projected position, line-of-sight velocity and proper motion of the stars. A comprehensive dynamical analysis of the data has inferred upper limits on the mass of IMBHs and led to several tentative detections \citep{2000AJ....119.1268G,2002AJ....124.3255V,2006ApJS..166..249M,2007ApJ...668L.139L,2010ApJ...710.1063V}.

The second line of evidence relies on observations in the radio and X-ray region. Numerous investigations focused on identifying IMBHs within GCs by analyzing their distinctive radiation signatures during the accretion phase \citep{2002ApJ...578L..41G,2004MNRAS.351.1049M,2007ApJ...661L.151U,2008AJ....135..182B}. However, the gas and dust within GCs are extremely sparse \citep{1991Natur.352..221S,2001ApJ...557L.105F}, which may cause a very faint accretion signature. Consequently, only a few debatable findings have been acquired \citep{2007ApJ...661L.151U,2008MNRAS.389..379M,2011ApJ...729L..25L,2018ApJ...862...16T,2021MNRAS.503.1490H}. A convincing evidence came from a luminous X-ray outburst in a massive star cluster, resulting from a tidal disruption event and subsequent accretion. The analysis of the spectra and light curve from this event demonstrates that the extragalactic cluster hosts an IMBH with a mass of $\sim 10^4~M_{\odot}$ \citep{2018NatAs...2..656L}.

The third piece of evidence comes from pulsar timing observations. To date, 305 pulsars have been discovered in 40 Galactic GCs\footnote{https://www3.mpifr-bonn.mpg.de/staff/pfreire/GCpsr.html for an up-to-date list}, with the overwhelming majority being millisecond pulsars (MSPs). This number continues to increase steadily. Pulsars are powerful tools to probe the GC dynamics \citep{1987MNRAS.225P..51B,1992RSPTA.341...39P}. Owing to their remarkably stable spin periods, measurements of pulsar acceleration and the local GC potential are feasible through the Doppler effect \citep{1993ASPC...50..141P,2017MNRAS.468.2114P,2017Natur.542..203K,2017MNRAS.471..857F,2017ApJ...845..148P,2019ApJ...884L...9A}. The high-eccentricity orbit of PSR B1820-30A indicates that an IMBH with a mass $>7500~M_{\odot}$ could potentially be located at the center of NGC 6624 \citep{2017MNRAS.468.2114P}. The measurement of the gravitational potential in the GC 47 Tucanae, using pulsars as indicators, offers evidence for a central IMBH with an estimated mass of $2300^{+1500}_{-850}~M_{\odot}$ \citep{2017Natur.542..203K}. By analyzing the measured accelerations of the pulsars in NGC 6266, \cite{2019ApJ...884L...9A} found that the center of the cluster displays an excess of dark mass ranging from $1200$ to $6000~M_{\odot}$. However, from a modeling perspective, it is worth noting that the central concentration of the stellar mass dark remnants may also account for all of the IMBH signatures identified in these investigations.

In this study, we concentrate on the GC NGC 6517, which is a metal-poor bulge cluster situated approximately $4.2~{\rm kpc}$ from the Galactic Center and around $10.6~{\rm kpc}$ from the Sun \citep{1996AJ....112.1487H}. So far, there is no positive report on the presence of an IMBH in this cluster. This includes all available data from optical, X-ray, and radio observations, as well as the pulsar timing method. NGC 6517 contains 15 confirmed radio pulsars, 9 of which have both period and period time derivative measurements \citep{2011ApJ...734...89L,2021RAA....21..143P,2021ApJ...915L..28P}. Five pulsars are located very close to the center of the GC ($\lesssim 0.2~{\rm pc}$), which might be considerably influenced by the possible IMBH in the center. We perform simulations to analyze the acceleration distributions of pulsars within the cluster, by utilizing optical observation data for the core radius and central density of the cluster \citep{1996AJ....112.1487H,2018MNRAS.478.1520B}. We compare the simulated results with the measured accelerations of the nine pulsars, which can offer a lower limit on the mass for the IMBH or central dark remnants. In Section 2, we outline the simulation methods, followed by the presentation of fitting results in Section 3. Finally, a summary of our findings can be found in Section 4.

\section{Methods}
%We establish a coordinate system for determining the positions of pulsars in the GC NGC 6517, as illustrated in Figure \ref{fig:1}. Define plane $O$ as the one perpendicular to the line-of-sight and passing through the center of gravity (COG) of the cluster. The variable $l$ represents the separation along the line-of-sight direction between the pulsar and plane $O$. The projected distance on the plane of the sky from the pulsar to the COG is denoted by $R_{\perp}^{\prime}$. Lastly, $r^{\prime}$ refers to the radial distance of the pulsar from the COG.

\begin{figure}
%\hspace{-0.5cm}
\centering
\includegraphics[width=0.7\columnwidth]{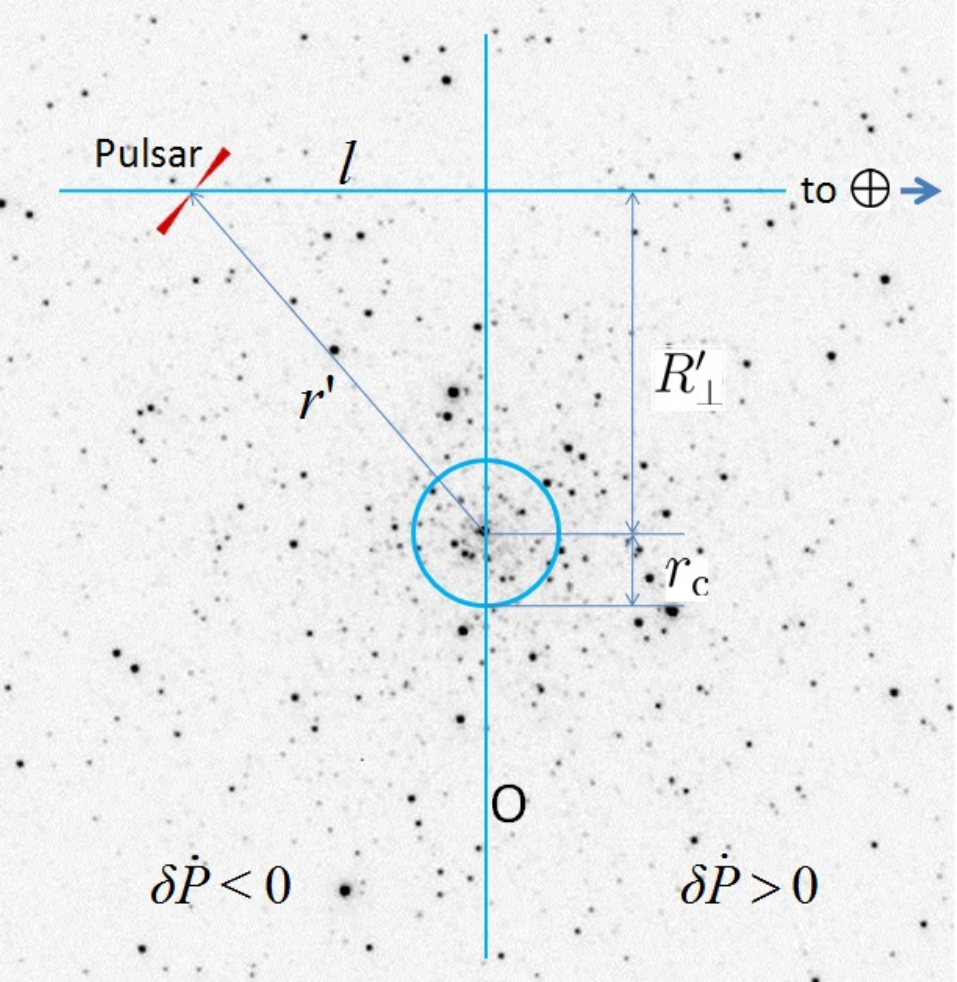}
\caption{Schematic illustration indicating the placement of a pulsar in relation to the central region of the cluster NGC 6517. The optical image of the cluster comes from the \emph{Hubble Space Telescope} (https://hla.stsci.edu/). The plane intersecting the cluster's COG and standing perpendicular to our line-of-sight is defined as plane $O$. The pulsar's line-of-sight distance from this plane is indicated as $l$. The symbol $R_{\perp}^{\prime}$ represents the measurement of the pulsar's projected distance from the cluster's gravitational center, while $r^{\prime}$ signifies the radial distance between them. Core radius is denoted as $r_{c}$. Accordingly, when the pulsar settles into the foremost half of the cluster, the GC potential induces a positive fluctuation in the period's time derivative. In contrast, if it positions on the rear half, it will contribute to a negative value.}
\label{fig:1}
\end{figure}

In this study, a coordinate system is proposed to determine the positions of pulsars in the GC NGC 6517, as shown in Figure \ref{fig:1}. The plane denoted as $O$ is defined as the one perpendicular to the line-of-sight and passing through the center of gravity (COG) of the cluster. The separation along the line-of-sight direction between the pulsar and plane $O$ is represented by the variable $l$. The projected distance on the plane of the sky from the pulsar to the COG is indicated as $R_{\perp}^{\prime}$. Finally, the radial distance of the pulsar from the COG is referred to as $r^{\prime}$.

The gravitational acceleration experienced by a test star, specifically a pulsar, within GCs stems from the combined influence (mean field) of all other stars contained in the cluster. The density distribution of these stars, extending from the center to a few core radii of the GC, can be formulated as King density profile \citep{1962AJ.....67..471K}:
\begin{equation}\label{King Model}
\rho(r^{\prime})=\rho_{\rm c}[1+(r^{\prime}/r_{\rm c})^2]^{-\frac{3}{2}},
\end{equation}
where $\rho_{\rm c}$ is the core density of the cluster, its core radius is $r_{\rm c}$.

When an IMBH is located at the center of a cluster, it significantly alters the spatial distribution of stars surrounding it. The IMBH with a specified mass $M_{\rm BH}$, has an associated influence radius \citep{2004ApJ...613.1133B}
\begin{equation}\label{influence radius}
r_{i}=\frac{3M_{\rm BH}}{8\pi \rho_{c}r_{c}^2}.
\end{equation}
The standard King model governs the star distribution at and beyond $r_i$ (up to a few core radii). However, within $r_{i}$, it is the density profile that complies with the formula \citep{2004ApJ...613.1143B},
\begin{equation}\label{power law}
\rho_{\rm BH}\propto r^{-1.55}.
\end{equation}
This power law distribution is deemed most suitable when modelling cusp clusters \citep{1986ApJ...305L..61D}.

The gravitational attraction felt by a pulsar at radius $r^{\prime}_{\ast}$ is generated by the stars enclosed within the spherical volume of radius $r^{\prime}_{\ast}$. By integrating Equation \ref{King Model} and \ref{power law} in radial direction, we can determine the total interior mass inside the sphere of radius $r^{\prime}_{\ast}$. Multiplying $G/r^{\prime 2}_{\ast}$ by the total interior mass yields the accerelation \citep{2020RAA....20..191X,2021RAA....21..270W,2022RAA....22k5007W}
\begin{equation}\label{Acceleration}
a(r^{\prime}_{\ast})=\frac{4\pi G}{r^{\prime 2}_{\ast}}[M_{\rm BH}/4\pi+\int_{0}^{r_{i}}r^{\prime 2}\rho_{\rm BH}dr^{\prime}+\int_{r_{i}}^{r^{\prime}_{\ast}}r^{\prime 2}\rho(r^{\prime})dr^{\prime}],
\end{equation}
in which $G$ is the gravitational constant. The mean field line-of-sight acceleration $a_{l}$, can be calculated by multiplying $a(r^{\prime}_{\ast})$ with a factor of $l/r^{\prime}_{\ast}$.

To investigate the distributions of $a_{l}$ with relation to $R_{\perp}^{\prime}$ for pulsars within the central region of the GC, we accomplish a thorough distribution profile through Monte Carlo simulations. Pulsars, being among the heavier stellar populations, are predominantly located in the innermost sector of the cluster due to mass segregation effects \citep{1987degc.book.....S}. This occurrence is well endorsed by notable observations made within the GCs 47 Tucanae and Terzan 5 \citep{2017MNRAS.471..857F,2017ApJ...845..148P}. In the observed scenario, the column number density profile associated with these deposited pulsars fits the description provided by \citep{1995ApJ...439..191L}
\begin{equation}\label{density profile}
n(x_{\perp})=n_0(1+x_{\perp}^2)^{\beta/2},
\end{equation}
where the central number density is denoted by $n_0$, the projected distance between the pulsar and the COG is given by $x_{\perp}\equiv R_{\perp}^{\prime}/r_{c}$. The mass segregation is described by the parameter $\beta$, which follows a Gaussian distribution centered at $-3$ with a dispersion of $0.5$ \citep{2019ApJ...884L...9A,2021RAA....21..270W}.

The time derivative of the pulsar spin period can be greatly influenced by the line-of-sight accelerations as a result of the Doppler effect. The first derivative of period $\dot P$ relates with the various accelerations in the following way \citep{1993ASPC...50..141P,2017ApJ...845..148P,2021RAA....21..270W,2022RAA....22k5007W}:
\begin{equation}\label{pdot}
\frac{\dot P}{P}=\frac{\dot P_0}{P_0}+\frac{a_{l}}{c}+\frac{a_{\rm g}}{c}+\frac{a_{\rm s}}{c}+\frac{a_{\rm DM}}{c}
\end{equation}
in which $\dot P_0/P_0$ is the intrinsic spin-down caused by the magnetic braking of a pulsar, $a_l$ is the line-of-sight acceleration due to GC potential, $a_{\rm g}$ is the acceleration from the Galactic potential, $a_{\rm s}$ accounts the centrifugal acceleration due to the Shklovskii effect, $a_{\rm DM}$ is the apparent acceleration due to the measuring error from the changes of dispersion measure (DM), and $c$ is the speed of light.

For a standard dipole radiation model, the intrinsic period derivative for a typical MSP can be estimated as
\begin{equation}\label{pdot0}
\frac{\dot P_0}{P_0}=3.1\times 10^{-10}~{\rm yr}^{-1}(\frac{B}{2\times 10^8~{\rm G}})^2(\frac{2~{\rm ms}}{P})^2
\end{equation}
where $B$ is the magnetic field strength at the surface of the pulsar.

By substituting typical values for GCs such as NGC 6517, and considering the case without an IMBH at the core, we can calculate the acceleration $a_{l}$ using Equation \ref{Acceleration}, which is expressed as follows  \citep{2005ApJ...621..959F,2017ApJ...845..148P}:
\begin{equation}\label{al}
\begin{split}
\frac{a_{l}}{c}=-3.7\times 10^{-9}~{\rm yr}^{-1} (\frac{\rho_{c}}{10^{5}~M_{\odot}~{\rm pc}^{-3}})(\frac{l}{0.2~{\rm pc}})(\frac{r}{r_{c}})^{-3} \\ \times[\sinh^{-1}(\frac{r}{r_{c}})-\frac{r}{r_{c}\sqrt{1+(r/r_{c})^2}}].
\end{split}
\end{equation}
The acceleration due to the Galactic potential can be derived from the pattern of Galactic rotation \citep{1995ApJ...441..429N}. We assume a flat rotation curve and use a distance for the Sun from the Galactic center $R_0=8.34\pm 0.16~{\rm kpc}$ and rotation velocity $\Theta_0=240\pm8~{\rm km~s^{-1}}$ \citep{2014ApJ...793...51S}. The line-of-sight acceleration between the Sun and each pulsar is given by \citep{1995ApJ...441..429N}
\begin{equation}\label{ag}
\boldsymbol{a_{\rm g}}\cdot\boldsymbol{n}=-\cos(b)(\frac{\Theta_0^2}{R_0})[\cos(l)+\frac{\beta}{\sin^2(l)+\beta^2}],
\end{equation}
where $\beta=(d/R_0)\cos(b)-\cos(l)$, and the distance of NGC 6517 from the Sun is $d=10.6~{\rm kpc}$. The Galactic coordinates at COG of the GC are $l=19^\circ.23$ and $b=6^\circ.76$. Then we find the acceleration is $a_{\rm g}/c=-5.75\times10^{-11}~{\rm yr}^{-1}$.

The Shklovskii effect is a centrifugal acceleration due to proper motion of the pulsar, which reads \citep{1970SvA....13..562S}
\begin{equation}\label{Shklovskii}
\frac{a_{\rm s}}{c}=8.1\times 10^{-13}~{\rm yr}^{-1} (\frac{d}{10.6~{\rm kpc}})(\frac{\mu_T}{{\rm mas~yr^{-1}}})^2.
\end{equation}
As NGC 6517 is situated close to the Galactic center, the proper motion $\mu_T$ may be estimated as \citep{2017ApJ...845..148P}
\begin{equation}\label{Proper motion}
\mu_T=4.7~{\rm mas~yr^{-1}}(\frac{\Theta_0}{240~{\rm km~s^{-1}}})(\frac{10.6~{\rm kpc}}{d}).
\end{equation}
The value is in good agreement with the precise proper motion of the cluster as measured in \cite{2021MNRAS.505.5978V}, which gives $\mu_T\simeq4.73\pm 0.03~{\rm mas~yr^{-1}}$. Using Equation \ref{Shklovskii}, we find the value $a_{\rm s}/c\sim 1.8\times 10^{-11}~{\rm yr}^{-1}$.

%\begin{table*}
\begin{table*}
\begin{center}
%\begin{minipage}[]{100mm}
\caption[]{List of Pulsars with Reported Period and its Time Derivative in NGC 6517\label{Tab:1}}%\end{minipage}
%\setlength{\tabcolsep}{2.5pt}
%\small
\begin{tabular}{c|cccccccc}
\hline \hline

Pulsar & $P$    & $\dot P$ & $\dot P/ P$ & $\theta_{\perp}$ & $R_{\perp}^{\prime}(1)$ & $R_{\perp}^{\prime}(2)$ & $R_{\perp}^{\prime}(3)$ & References\\
       & (ms)   & (10$^{-19})$ & ($10^{-9}$~yr$^{-1}$) & (arcmin) & (pc)&(pc)&(pc)&   \\
\hline
A  & 7.17  & -5.13   & -2.250 &0.022 & 0.068  & $0.059\pm 0.004$  & $0.065\pm 0.011$  &   1   \\
B  & 28.96 & 21.91   & 2.368  &0.022 & 0.068  & $0.059\pm 0.004$  & $0.065\pm 0.011$  &   1  \\
C  & 3.74  & -0.65   & -0.548 &0.057 & 0.176  & $0.153\pm 0.009$  & $0.169\pm 0.028$  &   1  \\
D  & 4.23  & 0.07    & 0.051  &1.202 & 3.706  & $3.227\pm 0.196$  & $3.566\pm 0.595$  &   1  \\
E  & 7.60  & -10.37  & -4.297 &0.025 & 0.078  & $0.068\pm 0.004$  & $0.075\pm 0.012$  &   2  \\
F  & 24.89 & -26.72  & -3.385 &0.055 & 0.168  & $0.147\pm 0.008$  & $0.162\pm 0.027$  &   2  \\
G  & 51.59 & 0.80    & 0.049  &0.121 & 0.373  & $0.325\pm 0.020$  & $0.359\pm 0.060$  &   2  \\
H  & 5.64  & 0.34    & 0.192  &0.562 & 1.734  & $1.509\pm 0.091$  & $1.667\pm 0.277$  &   3  \\
I  & 3.25  & -0.14   & -0.139 &0.757 & 2.335  & $2.032\pm 0.124$  & $2.246\pm 0.374$  &   4  \\

\hline \hline
\end{tabular}
\end{center}
%$^\ast$
Notes.  $\theta_{\perp}$ is the angular offset from the centre of the cluster for the pulsars in NGC 6517. The projected distances, denoted as $R_{\perp}^{\prime}(1)$, $R_{\perp}^{\prime}(2)$, and $R_{\perp}^{\prime}(3)$, are computed using $\theta_{\perp}$ and the distinct measurements of the distance of the cluster from the Sun. Each symbol correlates with specific distances: $d=10.6~{\rm kpc}$, $9.23\pm 0.56~{\rm kpc}$, and $10.2\pm 1.7~{\rm kpc}$, respectively.

\emph{References:} (1) \cite{2011ApJ...734...89L}; (2) \cite{2021RAA....21..143P}; (3) \cite{2021ApJ...915L..28P}; (4) Yin et al. 2023 (in preparation).

\end{table*}
%\end{table*}

The apparent acceleration caused by the stochastic DM error can be expressed as \citep{2017ApJ...845..148P}
\begin{equation}\label{aDM}
\frac{a_{\rm DM}}{c}=-6.4\times 10^{-15}~{\rm yr}^{-1}(\frac{\Delta t_{\rm DM}}{1 ~\rm{\mu s}})(\frac{10~\rm yr}{T})^2
\end{equation}
where $\Delta t_{\rm DM}$ is the delay time of the dispersion, $T$ is the timescale for the DM error measurement. For typical values of $\Delta t_{\rm DM}=100~\rm{\mu s}$ and $T=10~\rm yr$, one can find the apparent acceleration is $a_{\rm DM}/c=-6.4\times10^{-13}~\rm{yr}^{-1}$.

\begin{figure}
%\hspace{-0.5cm}%
\centering
\includegraphics[width=1.0\columnwidth]{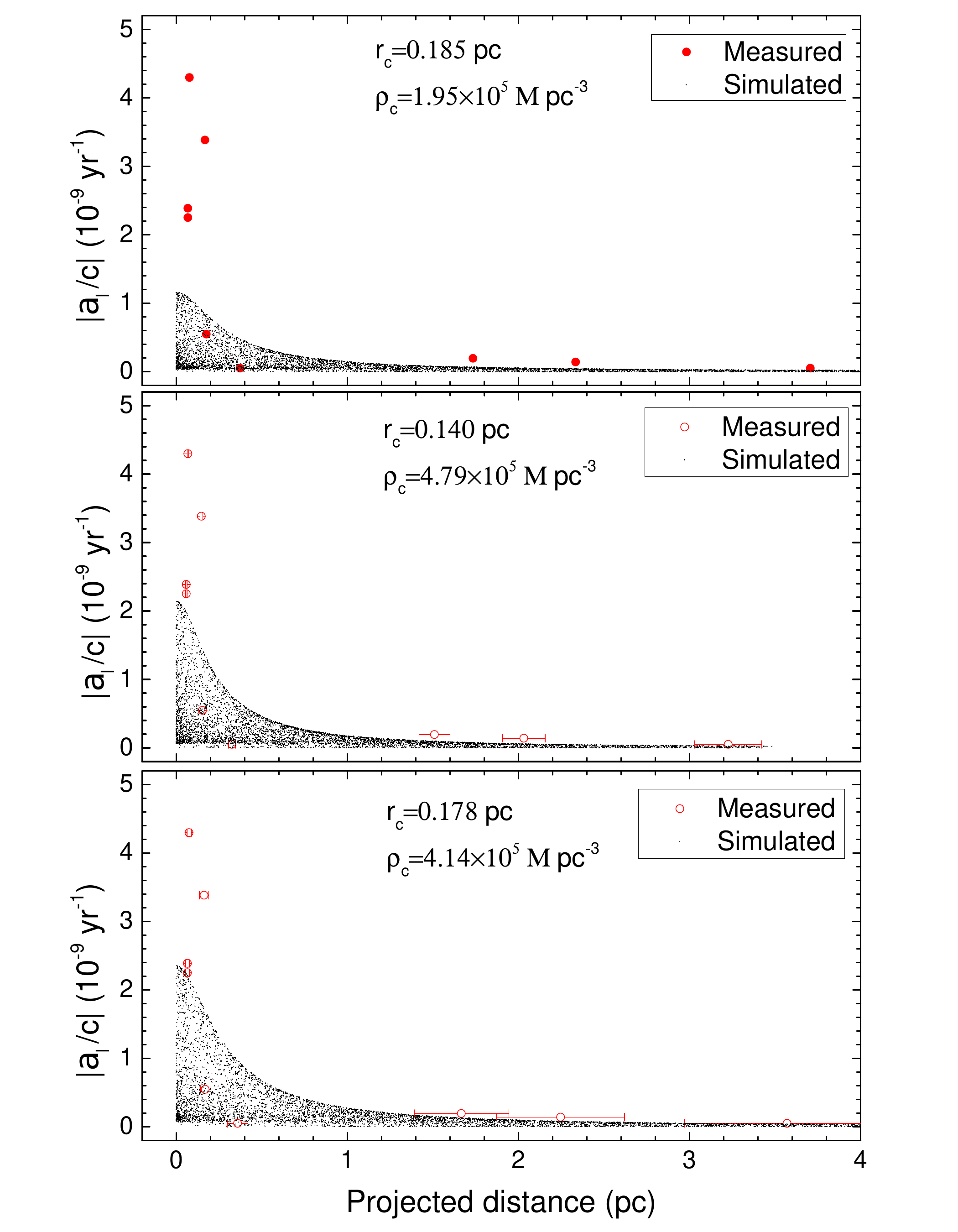}
\caption{The comparison of the measured $a_l$ with the simulated values, for pulsars in the innermost region of NGC 6517. Upper panel: $d=10.6~{\rm kpc}$, $r_{c}=0.185~{\rm pc}$, $\rho_{c}=1.95\times 10^5~M_{\odot}{\rm pc}^{-3}$ are taken from optical observation of Harris (1996). Middle panel: $d=9.23\pm 0.56~{\rm kpc}$, $r_{c}=0.14~{\rm pc}$, $\rho_{c}=4.79\times 10^5~M_{\odot}{\rm pc}^{-3}$ are taken from Baumgardt's catalog. Bottom panel: the distance from Gaia data, $d=10.2\pm 1.7~{\rm kpc}$, and the core radius $r_{c}=0.178~{\rm pc}$, and the central density derived from central velocity dispersion of the cluster $\rho_{c}=4.14\times 10^5~M_{\odot}{\rm pc}^{-3}$ are assumed. The central IMBH is not included in the simulations.}
\label{fig:2}
\end{figure}

It is evident that the terms of $a_{\rm g}$, $a_{\rm s}$ and $a_{\rm DM}$ contribute minimally to the measured acceleration of each pulsar. Consequently, Equation \ref{pdot} can be simplified to \citep{2022RAA....22k5007W}
\begin{equation}\label{pdot1}
\frac{\dot P}{P}=\frac{\dot P_0}{P_0}+\frac{a_{l}}{c}.
\end{equation}
In the Australian Telescope National Facility (ATNF) catalog \footnote{https://www.atnf.csiro.au/research/pulsar/psrcat/} \citep{2005AJ....129.1993M}, when we examine the MSPs, we find that $\dot P_0$ varies between $10^{-21}$ and $10^{-19}$ for $P_0<10~{\rm ms}$, resulting in a maximum value of $\dot P0/P0\sim 3\times 10^{-9}~{\rm yr^{-1}}$. This value is of the same order as the $a_{l}$ term. Due to the difficulty in distinguishing the contribution of the intrinsic spin-down term from the measured accelerations, this issue could introduce significant uncertainties in any research focused on measuring the GC potential.

\begin{figure*}
\centering
\includegraphics[width=0.50\textwidth]{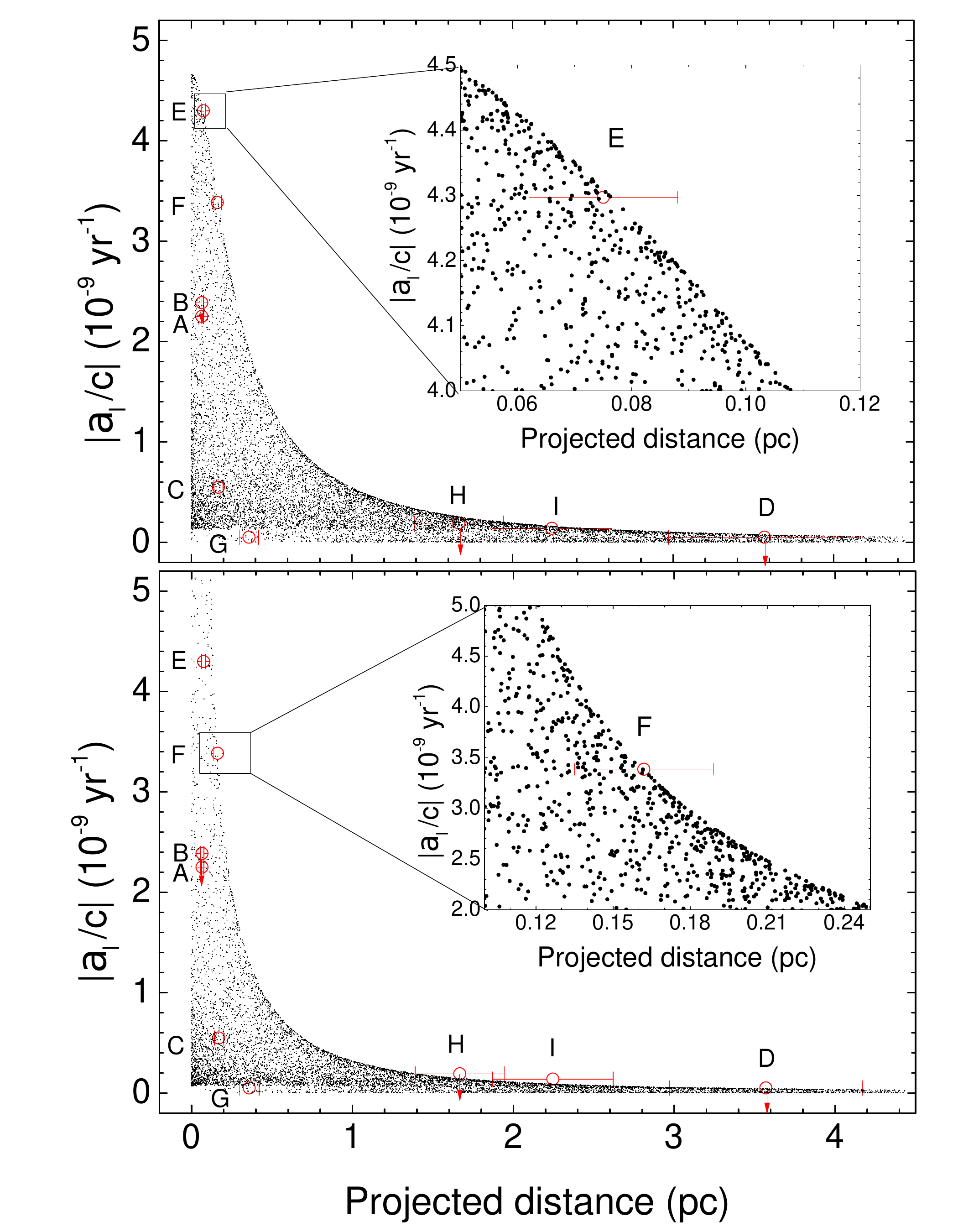}
\caption{The line-of-sight acceleration $a_l$ with respect to the projected distance $R_{\perp}^{\prime}$. The measured $a_l$ for pulsars from J1801-0857A to I are represented by red points and marked appropriately. All the pulsars with positive $\dot P$ are indicated with red arrows. The simulated results are caused by the gravitational potential of NGC 6517. The distance and core radius in the simulations are maintained at constant values of $d=10.2\pm 1.7~{\rm kpc}$ and $r_{c}=0.178~{\rm pc}$, respectively. Upper panel: the core density $\rho_{c}=8.2 \times 10^5~M_{\odot}{\rm pc}^{-3}$ is adopted. The central IMBH is not included in the simulation. Bottom panel: the core density derived from the central velocity dispersion $\rho_{c}=4.14\times 10^5~M_{\odot}{\rm pc}^{-3}$ is taken, and the central IMBH has a mass of $M_{\rm BH}=9000~M_{\odot}$ is assumed.}\label{fig:3}
\end{figure*}

It is important to note that $\dot P_0$ is always positive. In cases where a pulsar is situated in the rear half of the cluster, the gravitational potential of the GC negatively impacts the time derivative of the period (i.e., $\delta\dot P<0$). This is illustrated in Figure \ref{fig:1}. Consequently, pulsars with a negative $\dot P$ are likely influenced by the GC potential as the dominant term, represented by $\frac{\dot P}{P}\approx \frac{a_{l}}{c}$. This relationship can be utilized to provide a minimum estimate for the GC potential and the mass of a possible IMBH.

\section{Results}
\label{sec:analysis}

In NGC 6517, nine pulsars with measured $P$ and $\dot P$ have been observed, as shown in Table \ref{Tab:1}. Among them, five pulsars, J1801-0857A, C, E, F and I, have negative $\dot P$, which makes the cluster a very good candidate for gravitational field estimation.

In Table \ref{Tab:1}, we display the angular offsets $\theta_{\perp}$ from the centre of the cluster for the nine pulsars. The projected distances for pulsars, can be determined by using the offsets $\theta_{\perp}$ and the distance $d$ of the cluster from the Sun. We utilize three distinct measurements of the distance $d$ in determining the projected distances: the distance taken from the optical observation of \cite{1996AJ....112.1487H} is $10.6~{\rm kpc}$ (without an error bar), the mean distance given by Baumgardt's list of fundamental globular cluster parameters (Baumgardt's catalog) \footnote{https://people.smp.uq.edu.au/HolgerBaumgardt/globular/parameter.html} is $9.23\pm 0.56~{\rm kpc}$, and the recent value derived from Gaia data is $10.2\pm 1.7~{\rm kpc}$ \citep{2021MNRAS.505.5957B}. Employing these specified measurements, we derived the respective projected distances, represented as $R_{\perp}^{\prime}(1)$, $R_{\perp}^{\prime}(2)$, and $R_{\perp}^{\prime}(3)$, which have been displayed in the same table.

We calculate the line-of-sight acceleration $a_l$ for pulsars located at various radial distances in NGC 6517 with Equations \ref{King Model}-\ref{density profile}. We perform Monte Carlo simulations on the acceleration $a_l$ with respect to the projected distance, for pulsars in the innermost region of the cluster. The simulated $a_l$ distributions are shown in Figure \ref{fig:2}. At first, we adopt the cluster distance from the Sun $d=10.6~{\rm kpc}$, in association with the core radius $r_{c}=0.185~{\rm pc}$, which is derived from the angular core radius $\theta_{\rm c}=0.06~{\rm arcmin}$, and core density $\rho_{c}=1.95\times 10^5~M_{\odot}{\rm pc}^{-3}$, as determined from the optical observations for NGC 6517 \citep{1996AJ....112.1487H}, and the simulated results are shown in the upper panel of Figure \ref{fig:2}. When compared to the measured accelerations, the simulated results are noticeably smaller. This indicates that the central density $\rho_{c}$ from the optical observation is significantly lower than the mass density generating the GC potential.

We then take the following parameters from Baumgardt's catalog: the distance of $d=9.23\pm 0.56~{\rm kpc}$, the core radius of $r_{c}=0.14~{\rm pc}$, and the core density of $\rho_{c}=4.79\times 10^5~M_{\odot}{\rm pc}^{-3}$. The last parameter was estimated using N-body simulations as detailed in the same catalog. Applying these values, we proceed with the simulation, and show the simulated results in the middle panel of Figure \ref{fig:2}. The error bars of the measured values for the position of pulsars are derived from the distance $d$. One can see that the simulated distribution remains below the measured $a_l$ values for certain pulsars.

We utilize the distance from Gaia data, $d=10.2\pm 1.7~{\rm kpc}$, and the core radius $r_{c}=0.178~{\rm pc}$. The central density can be estimated by using the central velocity dispersion of the cluster, which is defined as \citep{1966AJ.....71...64K,2005ApJS..161..304M,2017MNRAS.471..857F}
\begin{equation}\label{dispersion}
\rho_{c}=\frac{9\sigma_0^2}{4\pi G r_{c}^2},
\end{equation}
where $\sigma_0$ is the central velocity dispersion, and for NGC 6517, $\sigma_0=9.2~{\rm km/s}$ is taken from Baumgardt's catalog. Then we obtain $\rho_{c}=4.14\times 10^5~M_{\odot}{\rm pc}^{-3}$. The bottom panel of Figure \ref{fig:2} shows the simulated results. It is evident that, for specific pulsars, the simulated distribution persistently falls short of the measured $a_l$ values.

By retaining $d=10.2\pm 1.7~{\rm kpc}$ and $r_{c}=0.178~{\rm pc}$, and adjusting the mass density to $\rho_{c}=8.2^{+0.2}_{-0.2}\times 10^5~M_{\odot}{\rm pc}^{-3}$, we find that all the measured values are encompassed by the simulated outcomes, as shown in upper panel of Figure \ref{fig:3}. The uncertainty arises due to the adjustment process with respect to the error bar of $R_{\perp}^{\prime}(3)$ for Pulsar J1801-0857E. This strongly implies that some optically nonluminous matters are necessary for explaining the measured accelerations of the pulsars. Potential candidates include massive white dwarfs, neutrons stars or stellar mass BHs that formed in the early phase of the GC. In addition to a concentration of dark remnants in the central region, it is also plausible that an IMBH exists at the core of the GC.

For the models, we consider the presence of an IMBH at the center of the GC. We perform the simulations using various masses of the IMBH, while maintaining $d=10.2\pm 1.7~{\rm kpc}$, $r_{c}=0.178~{\rm pc}$. We adopt the core density derived from the central velocity dispersion, $\rho_{c}=4.14\times 10^5~M_{\odot}{\rm pc}^{-3}$. This value is in approximate consistency with results obtained from N-body simulations of Baumgardt's catalog. The results of the simulations can be found in the bottom panel of Figure \ref{fig:3}. The pulsars J1801-0857E and F, exhibiting the highest measured accelerations, can be well covered with the simulated data for the mass of the IMBH $M_{\rm BH}=9000^{+4000}_{-3000}~M_{\odot}$. The uncertainty emerges as a consequence of the adjustment process with respect to the error bar of Pulsar J1801-0857F. The analysis also reveals that the simulated data cannot account for the measured acceleration of pulsar I. However, density model used for the entire cluster (Equation \ref{King Model}) describes the density profile only within a few core radii of a GC. It is found from Equations (14) and (27) reported in \cite{1962AJ.....67..471K} setting the tidal radius to infinity. It is not a good approximation of the actual density far from the centre of the GC. For this reason, the fitting should not be applied to pulsars H, I and D without substantial justifications.

It should be pointed out that an IMBH of the mass $M_{\rm BH}\sim 9000~M_{\odot}$ would have an influence radius of $0.08~{\rm pc}$ ($\sim 1.7$ arcseconds), about $46\%$ of the core radius of the cluster. This would have strong effects on central stars of the GC and the radial profile would show a clear cusp in the central region. However, such a cusp is not seen in the surface density profiles of the cluster (see for example \cite{1995AJ....109..218T}). This absence suggests that the dark remnants model could be a more plausible interpretation. However, currently determining this is challenging. This is due to the fact that the segregation of stellar mass BHs, neutron stars and massive white dwarfs induces a steep increase in dark mass, subsequently causing a softening effect on the luminosity cusp. Furthermore, the irregular movement of an IMBH, when positioned relative to the cluster centre, may cause additional distortion, thereby further lessening the ascent of a centrally rising cusp \citep{2017Natur.542..203K}.

We would like to emphasize that the simulated $a_l$ distribution profiles for the dark remnants model (upper panel of Figure \ref{fig:3}) and the IMBH model (bottom panel of Figure \ref{fig:3}) exhibit considerable differences. These distinct distribution profiles could potentially be employed to differentiate between the two scenarios, once a sufficient number of pulsars have been discovered within the cluster.

%\& Discussions

\section{Summary}
\label{sec:discussion}
In NGC 6517, there are five pulsars exhibiting negative $\dot P$, rendering the cluster highly suitable for gravitational potential estimation. Utilizing the measured accelerations of these pulsars, we have predicted the gravitational potential and mass density of the cluster. Moreover, we conducted a Monte Carlo simulation on the acceleration distributions for pulsars, and compared them to the measured values. Our findings indicate a concentration of non-luminous mass in the central region of the cluster, and the source of this dark mass excess could be either a system of stellar dark remnants or a single IMBH with a mass of approximately $\gtrsim 9 \times 10^3~M_{\odot}$. Although our analysis cannot currently distinguish between these two scenarios, the simulations predict distinct acceleration distribution profiles for each. As a result, it may be possible to differentiate between the two once enough pulsars with negative $\dot P$ are discovered within the cluster.

% The two source have different $a_l$ distribtutions, we need more data to discriminate the two cases.

\section*{Acknowledgements}

We appreciate valuable discussions with C.Cheng and J.Y.Liao. We would like to thank the referee for comments that led to a significant improvement of this paper. ZP is supported by National Key R\&D Program of China, No. 2022YFC2205202, and the CAS ``Light of West China'' Program. This work is supported by National Natural Science Foundation of China under grant Nos. 11803009, 11603009, 11703047, 11773041, U2031119, and 12173052, and by the Natural Science Foundation of Fujian Province under grant Nos. 2023J01806, 2018J05006, and 2018J01416.

%%%%%%%%%%%%%%%%%%%%%%%%%%%%%%%%%%%%%%%%%%%%%%%%%%
\section*{Data Availability}

Data generated in this research will be shared on reasonable request to the corresponding author.



\bibliographystyle{mnras}
\bibliography{bibliography} % if your bibtex file is called example.bib

%\begin{thebibliography}{99}
%\end{thebibliography}

%%%%%%%%%%%%%%%%%%%%%%%%%%%%%%%%%%%%%%%%%%%%%%%%%%

% Don't change these lines
\bsp	% typesetting comment
\label{lastpage}
\end{document}